\documentclass[preprint,12pt]{elsarticle}

\usepackage{url}
\usepackage{tikz}
\newcommand*\circled[1]{\tikz[baseline=(char.base)]{
            \node[shape=circle,draw,inner sep=0.5pt] (char) {\scriptsize #1};}}

\usepackage[hidelinks]{hyperref} 
\usepackage{multirow} 
\usepackage{pifont}
\newcommand{\cmark}{\ding{51}}%
\newcommand{\xmark}{\ding{55}}%
\usepackage{footnote}
\usepackage{amsmath}
\usepackage{amssymb}
\usepackage{color}
\usepackage{soul}
\usepackage{amsmath} 
\usepackage{amsfonts} 
\usepackage{amssymb} 
\usepackage[ruled,vlined]{algorithm2e} 

\usepackage{makecell}

\usepackage{url}
\usepackage{breakurl}  
\usepackage{tablefootnote}
\usepackage[flushleft]{threeparttable}

\journal{Internet of Things}

\begin{document}

\begin{frontmatter}



\title{Towards Robust Stability Prediction in Smart Grids: GAN-based Approach under Data Constraints and Adversarial Challenges}


\author[label1]{Emad Efatinasab}
\author[label2]{Alessandro Brighente}
\author[label2]{Denis Donadel}
\author[label2]{Mauro Conti}
\author[label1]{Mirco Rampazzo}
\affiliation[label1]{organization={University of Padova, Department of Information Engineering},
city={Padova}, country={Italy}}
\affiliation[label2]{organization={University of Padova, Department of Mathematics},
            city={Padova},
            country={Italy}}
\begin{abstract}
Smart grids are crucial for meeting rising energy demands driven by global population growth and urbanization. By integrating renewable energy sources, they enhance efficiency, reliability, and sustainability. However, ensuring their availability and security requires advanced operational control and safety measures. 
Although artificial intelligence and machine learning can help assess grid stability, challenges such as data scarcity and cybersecurity threats, particularly adversarial attacks, remain.
Data scarcity is a major issue, as obtaining real-world instances of grid instability requires significant expertise, resources, and time. Yet, these instances are critical for testing new research advancements and security mitigations.
This paper introduces a novel framework for detecting instability in smart grids using only stable data. It employs a Generative Adversarial Network (GAN) where the generator is designed not to produce near-realistic data but instead to generate Out-Of-Distribution (OOD) samples with respect to the stable class. These OOD samples represent unstable behavior, anomalies, or disturbances that deviate from the stable data distribution. By training exclusively on stable data and exposing the discriminator to OOD samples, our framework learns a robust decision boundary to distinguish stable conditions from any  unstable behavior, without requiring unstable data during training. Furthermore, we incorporate an adversarial training layer to enhance resilience against attacks.
Evaluated on a real-world dataset, our solution achieves up to 98.1\% accuracy in predicting grid stability and 98.9\% in detecting adversarial attacks. Implemented on a single-board computer, it enables real-time decision-making with an average response time of under 7ms.

\end{abstract}

\begin{graphicalabstract}
\end{graphicalabstract}

\begin{highlights}
\item Tackles smart grid stability detection while addressing data scarcity.

\item Proposes a novel GAN framework detects grid instability using only stable data.

\item Generator creates OOD samples beyond the stable class to train the discriminator.

\item Added adversarial training layer to classify attacks as instability instances.

\item Deployed on single-board computer with under 7 ms avg response time.





\end{highlights}


\end{frontmatter}



\section{Introduction}

The rapid growth of the global population, economy, and urban areas is expected to significantly increase energy demand, coinciding with the rise of renewable energy.
Traditionally, energy grids featured a unidirectional flow from producers to consumers. But the emergence of prosumers—entities that both consume and supply energy—demands a shift to bidirectional flow~\cite{breviglieri2021predicting}.
Smart grids offer a transformative solution, enhancing the efficiency, reliability, and sustainability of electrical networks through advanced technologies.
They enable modern electricity distribution with improved dependability, effectiveness, and two-way communication~\cite{muqeet2023state}.
As the energy landscape evolves, smart grids are key to integrating renewable energies like solar and wind, addressing variability, and supporting sustainability.
Their adaptability is crucial for managing new technologies and operational features, such as power collection timing and delivery capacity~\cite{efatinasab2024gangrid}.

Accurate forecasting of renewable energy generation is vital for stable and efficient power system operations, particularly given the inherent variability of sustainable sources~\cite{Jiao_2020}. Similarly, robust forecasting methods help preempt disruptions in balancing electricity supply and demand.
To tackle the challenges of fluctuating power grids, smart grid strategies have emerged, with a key focus on supply-demand balancing. A central approach is the demand response strategy, where consumers adjust electricity use in response to price changes, deviating from typical consumption patterns~\cite{ALBADI20081989,5930335,en13102559}.
A promising advancement in this area is Decentralized Smart Grid Control (DSGC), which integrates electricity prices to grid frequency---a measurable parameter for prosumers~\cite{schafer2015decentral}. Frequency rises during surpluses and falls during shortages~\cite{4282051}, enabling real-time pricing that prompts prosumers to adapt demand dynamically.
However, effective DSGC implementation faces several challenges, such as ensuring grid stability amid rapid price shifts, handling varied price sensitivities, and accommodating differences in participant response times~\cite{8587498}.

Grid instability can cause major disruptions to the electricity supply, affecting daily life and economic systems. When the power grid becomes unstable, it can lead to outages, damage electrical equipment, and pose safety risks. For example, voltage fluctuations may cause lights to flicker and potentially damage sensitive devices~\cite{voltfix_grid_instability}. 
In extreme situations, a localized outage triggered by grid instability can cascade into widespread blackouts~\cite{efatinasab2024gangrid}. A real-world example is the near-total blackout in Puerto Rico, where a fault in an underground cable left approximately 1.3 million people without power~\cite{politico_puerto_rico_blackout}.

Machine Learning (ML) and Artificial Intelligence (AI) have proven highly effective for stability prediction in decentralized smart grids, with several models achieving near-perfect accuracy in detecting unstable samples~\cite{breviglieri2021predicting,10.1145/3459104.3459160,9079864,9446196}.
However, a major challenge remains: the lack of real-world datasets containing both stable and unstable instances. This gap poses a major obstacle to advancing the field with AI, as such datasets are essential for training reliable models.
The scarcity arises from the fact that unstable behaviors---typically indicating system failures, malfunctions, or sudden demand fluctuations---are both rare and undesirable.
Collecting such data in a real-world scenario is difficult and risky, as inducing unstable behaviors could cause serious disruptions, equipment damage, or safety issues. Moreover, building high-quality datasets is time-consuming and often requires manual labeling~\cite{9058539}.

Beyond data scarcity, another major challenge in smart grid applications is ensuring the security of AI-driven systems. As smart grids increasingly rely on data-driven technologies, robust security measures are crucial to protect the confidentiality, integrity, and availability of energy infrastructure~\cite{efatinasab2024faultguard}. 
While AI and ML techniques enhance predictive capabilities and grid management efficiency, their vulnerability to adversarial attacks remains a critical and often overlooked issue in the literature~\cite{10659889}.
The extensive interconnection between devices and remote access points expands the attack surface, creating potential entry points for attackers to infiltrate the entire network. 
Several studies~\cite{9914610,10.1145/3447555.3464859} have examined the susceptibility of AI-based stability prediction systems in smart grids to such attacks. For example, Efatinasab et al.~\cite{efatinasab2024gangrid} demonstrated how an attacker can exploit grid stability prediction models by injecting adversarial data, causing the system to misclassify unstable conditions as stable. 
This threat is particularly severe during high-demand periods, such as extreme weather events~\cite{KE2016504}, when power systems are under stress and require corrective measures. 
If misleading data prevents appropriate responses---like load shedding or activating backup systems---overloads and failures in critical components may occur.
However, to the best of our knowledge, no existing work addresses stability prediction using only stable data while also ensuring robustness against adversarial attacks within a single integrated model.

\emph{Contributions.} In this paper, we propose \textbf{GAN-Stability}, a novel framework for stability prediction and adversarial attack detection. Our approach is trained using only one class (stable samples) from a two-class dataset for the stability prediction task. 
Our solution trains the discriminator of a Generative Adversarial Network (GAN) with three types of data: i) synthetic samples generated by the GAN generator that are potentially Out-Of-Distribution (OOD) relative to the stable class, ii) stable samples from a real-world dataset, and iii) adversarial samples crafted by attacking the stable data.
By generating OOD samples synthetically, our method alleviates the need to induce unstable behaviors in the smart grid for data collection. 
These generated samples may reflect unstable behavior, measurement anomalies, fault-induced disturbances, or other deviations from normal behavior.
Moreover, GAN-Stability integrates adversarial detection directly into the GAN framework, removing the need for a separate model. 
Our approach trains the discriminator to identify adversarial samples as part of the unstable class within the same process, simplifying the overall architecture and improving robustness.
We tested our framework on a dataset containing both stable and unstable data. Our solution achieves an accuracy of up to 0.981 in stability prediction,  even when trained exclusively on stable data. We also evaluate the robustness of our adversarial training approach against state-of-the-art attacks in both whitebox and greybox scenarios, achieving accuracies up to 0.989 in classifying attacks as unstable behaviors. While reducing the burden on dataset development, our solution eliminates the need for a separate Anomaly Detection System (ADS) to identify adversarial attacks. 

Our contributions can be summarized as follows.
\begin{itemize}
    \item We propose \textbf{GAN-Stability}, a framework for training stability prediction systems using exclusively stable instances while maintaining high accuracy. To the best of our knowledge, we are the first to develop such a model in the smart grid context.
    \item We enhanced the capabilities of our model through adversarial training, enabling GAN-Stability to classify adversarial attacks as unstable samples without the need for an external ADS.
    \item We evaluate our system on a widely used dataset, achieving an accuracy of up to 0.981 in stability prediction . Additionally, the integration of adversarial training allows the model to detect state-of-the-art adversarial attacks with an accuracy of up to 0.989.
    \item We compare our solution with state-of-the-art models that require both stable and unstable data for training. Our model demonstrated superior or comparable performances, despite the reduced data requirement for training (i.e., only stable data).
    \item We test \textbf{GAN-Stability} on an affordable microcomputer, showing reasonable training time requirements and an average response time of less than 7ms during testing.
    
    \item We make the code of our systems, attacks, and the dataset available at: 
    \url{https://github.com/emadef1/GAN-Stability}
\end{itemize}

\emph{Organization.}
The rest of the paper is organized as follows: Section~\ref{sec:RW} reviews existing stability prediction systems and their associated security implications. Section~\ref{sec:system_model_threat_model} introduces the system and threat model, reflecting real-world scenarios encountered during training and potential attacks. Section~\ref{sec:attacks} discusses the adversarial attack methodologies employed in this study.  Section~\ref{methodology} details the proposed stability prediction system and its methodology, while Section~\ref{Eval} evaluates our proposed system using different criteria. Section~\ref{limit} outlines the limitations and discusses potential areas for improvement. Finally, Section~\ref{conclusion} concludes the study with some final remarks.

\section{Related Works}\label{sec:RW}
In this section, we dig into the existing literature concerning stability prediction systems and their security implications. Specifically, we scrutinize established methodologies for stability prediction in Section~\ref{sec:stability}, while looking at the current landscape of attacks targeting these systems in Section~\ref{sec:adversey}.


\subsection{Smart Grid Stability Prediction using AI}\label{sec:stability}
The rise of distributed and renewable energy sources presents significant challenges in ensuring the stability of power grids. 
While researchers have taken various approaches in the past~\cite{ourahou2020review}, ML and AI are shown to be an efficient way to enhance smart grid functionality by facilitating intelligent decision-making and rapid responses to various dynamic scenarios~\cite{9084590}. 
Advanced AI techniques provide robust solutions for stability analysis and control in smart grids, gathering considerable interest and attention from both researchers and practitioners~\cite{SHI2020115733}.  

For instance, Aliyeva et al.~\cite{10488774} developed a hybrid DL model that combines Multilayer Perceptron (MLP) and Extreme Gradient Boosting (XGBoost) classifiers to forecast smart grid stability. 
Bashir et al.~\cite{https://doi.org/10.1002/2050-7038.12706} employed various state-of-the-art ML algorithms, such as Support Vector Machines (SVM), K-Nearest Neighbor (KNN), Logistic Regression, Naive Bayes, Neural Networks, and Decision Tree classifiers, to predict smart grid stability. 
Gorzałczany et al.~\cite{en13102559} approach the issue of smart grid stability prediction by utilizing a knowledge-based data-mining technique, particularly focusing on a fuzzy rule-based classifier. 
Furthermore, there is a growing emphasis on the utilization of Recurrent Neural Networks (RNNs) such as Long Short-Term Memory Network (LSTM) and Gated Recurrent Unit (GRU) in the literature~\cite{10.1145/3459104.3459160,9079864}. 
Zhang et al.~\cite{10.1145/3459104.3459160} introduce a power grid stability prediction model that relies on a Bi-directional LSTM with an attention mechanism. 
This model is capable of learning the function of various stability features and the interrelationships among these features. 

A novel Multidirectional LSTM technique has been introduced by~\cite{9079864} for predicting the stability of smart grid networks. 
Furthermore, Massaoudi et al.~\cite{9446196} propose a DL approach using bidirectional GRU for predicting smart grid stability. 
To automate the tuning process, this research utilizes the Simulated Annealing algorithm to optimize selected hyperparameters and improve the model's forecasting capability. 
Also, the utilization of Convolutional Neural Networks (CNNs) in stability prediction research within smart grids has been explored by various researchers~\cite{8486644,SHI2020114586}. 
While all these represent viable solutions, model training is always employing stable data together with unstable samples. However, the assumption of having unstable data is not always achievable in real-world settings, thus creating the need for alternative systems that rely only on stable data. 
\subsection{ML Adversarial Attacks}\label{sec:adversey}
Recent studies have highlighted the vulnerabilities of various ML methods to adversarial attacks, raising concerns about their impact on the security and reliability of power systems~\cite{9652053}. Nowadays, smart grids are employing AI for grid stability, and adversarial examples can significantly compromise the outcome of these systems. Additionally, findings from~\cite{AYGUL2024101012} demonstrate that during cyber-attacks, ML algorithms suffer a notable drop in performance, leading to a sharp decline in the accuracy of transient stability predictions compared to normal conditions.
Furthermore, Chen et al.~\cite{8587547} aim to address security issues associated with ML applications in power systems. They emphasize that most ML algorithms proposed for power systems are susceptible to adversarial examples—inputs intentionally crafted with malicious intent. 

The paper by Tian et al.~\cite{tian2022adversarial} investigates security concerns of neural network-based state estimation in smart grids, focusing on adversarial attacks and proposing an efficient method for executing these attacks. 
Sayghe et al.~\cite{9281719} investigate the impact of adversarial examples on the detection of False Data Injection Attacks (FDIAs) using DL algorithms. Their research examines the repercussions on MLP when exposed to two different adversarial attack strategies.
Ahmadian et al.~\cite{8646424} introduced a FDIA using a GAN framework, where the attacker acts as the generative network and the Energy System Operator (ESO) serves as the discriminative network. The attacker generates deceptive data to evade detection by the power system state estimator through an optimization process. 
Li et al.~\cite{9303013} show that well-established ML models used in energy theft detection systems are susceptible to adversarial attacks. They develop a method to create adversarial measurements, allowing attackers to report significantly lower power consumption to utility companies and evade detection by ML-based systems. In addition, Song et al.~\cite{10.1145/3447555.3464859} conducted a comprehensive analysis of adversarial example attacks in the context of voltage stability assessment for the New England 10-machine 39-bus system. Their study evaluated the reliability of six key attack methods, revealing that most could reduce the target deep neural network’s accuracy by approximately 50\% when modifying only half of the input dimensions.


\section{System and Threat Model}
\label{sec:system_model_threat_model}

In this section, we introduce the system and threat model for our GAN-Stability framework.

\subsection{System Model}
\label{system_model}


In an operational setting without active threats, a DSGC stability prediction system assesses whether the grid remains stable or unstable, particularly in a decentralized smart grid context where electricity prices are tied to grid frequency that carry all necessary information
about the current power balance.
In fact, the stability of electrical grids depends on the balance between electricity generation and  demand~\cite{en13102559,8587498}.
In the context of DSGC, stability is characterized by synchronized node frequencies (\(\omega\)) and steady power flows (\(P_{jk}\)) across the grid. Stability requires minimal angular frequency deviations and effective damping to suppress oscillations. It is evaluated by the system's ability to return to equilibrium after disturbances, as measured by mathematical models such as local stability (i.e., linear stability exploring dynamical stability around the steady-state operation of the grid)~\cite{8587498}. Non-stable behaviors arise when synchronization is lost, resulting in significant frequency deviations, amplified oscillations, and destabilized power flows. These effects can lead to cascading failures, particularly when delays, resonance effects, or insufficient damping prevent the grid from recovering.


Our system utilizes ML and AI algorithms to perform binary classification, categorizing grid samples into stable or unstable classes based on various input data collected by the control center from nodes on the grid.
Examples of such data include the reaction time of each participant, which indicates how quickly consumers or systems respond to changes; price elasticity coefficients, which reflect the sensitivity of power consumption to changes in electricity prices; and nominal power consumption and production features, which represent baseline levels of power used or generated by the system. 
As we will discuss in Section~\ref{subsec:arch}, our system employs 12 different features, typically collected by low-cost equipment from individual prosumers and sent to the control center.
Before deployment, the model is trained on clean, uncorrupted data to ensure reliable predictions. Stable grid instances are easy to obtain from operational data as they represent the majority of the system's operating time. In contrast, collecting unstable data is more challenging. Instability instances require careful labeling by human experts, and acquiring a comprehensive dataset demands long-term observation and significant resources.

In our system model, we adopt a pragmatic approach where we collect enough instances of stability from low-cost equipment by particular prosumers~\cite{en13102559}. These instances, all from the same stable label, serve as the sole data for training the stability prediction model. Therefore, to ensure the model's effectiveness, we should collect a sufficient amount of data that is both comprehensive and representative of the underlying distribution of stable grid conditions.
By focusing solely on stability instances, we streamline the training process and alleviate the need for extensive data collection efforts associated with capturing instances of instability. 
To the best of our knowledge, all the models in the existing literature typically rely on access to both labels in the dataset to make accurate predictions~\cite{breviglieri2021predicting,efatinasab2024gangrid,9079864,mohsen2023efficient,ALLAL2024108304,MOSTAFA2022100363}. 
Our approach challenges this conventional paradigm by demonstrating that accurate stability prediction can be achieved using only instances of stability for training. 

\subsection{Threat Model}\label{threat}
The attacker's objective is to stealthily insert fraudulent information into the grid's data stream, manipulating the classification decisions made by the stability model. This manipulation can result in misclassification in both directions---either causing stable grid conditions to be incorrectly classified as unstable or, more critically, unstable conditions to be classified as stable. 
In pursuit of this goal, the attacker may exploit either known vulnerabilities or discover new ones to gain remote access to the smart grid elements~\cite{sullivan2017cyber}. 
We delineate two scenarios based on the attacker's familiarity with the data of the smart grid and the stability prediction model.
\begin{itemize}
    \item \textit{White-box Scenario}: In this scenario, the attacker possesses comprehensive access to both the data employed in testing the model and detailed information regarding the model's architecture and parameters. 
    This advantageous position provides the attacker with ample opportunities to exploit vulnerabilities in the system. 
    By leveraging this intelligence, the attacker can meticulously craft powerful adversarial samples aimed at deceiving the model.
    Such a situation could arise if an attacker compromises the control center of the smart grid, for instance, through malware infiltration targeting electric power systems---such as Industroyer~\cite{10646775}, which was used in the Ukrainian power grid attack. Additionally, physical attacks on power companies~\cite{xu2020analysis} represent another real-world threat that could enable such an attack.
    
    \item \textit{Grey-box Scenario 1}: In this scenario, the attacker has access to the testing data but lacks access to the architecture and parameters of the main model. 
    Despite this limitation, the attacker can still conduct evasion attacks by employing a surrogate model---an alternative model trained on the same dataset---with different architectures. The effectiveness of these attacks relies on transferability properties or the chosen architecture. This condition can exist if an attacker compromises enough prosumers or entities in the system, thus gaining access to several data points but without knowledge of the actual employed model. IoT botnets are an example of how such scenarios could easily become a reality~\cite{ali2020systematic}. By compromising large numbers of devices, IoT botnets can collect data or probe the system, potentially providing the attacker with enough information to train a surrogate model. Additionally, attacks such as the Man-in-the-Middle (MitM) attack have been shown to disrupt communication between control systems and field equipment in smart grids. For instance, \cite{9640002} demonstrates how MitM attacks can leverage false data injection techniques to alter transmitted data, including issuing deceptive commands to field devices.
    
    To generate adversarial samples, we utilize the LSTM model proposed in~\cite{efatinasab2024gangrid} as a surrogate model. These adversarial samples will then be deployed against our primary stability prediction system (GAN-Stability). This setup simulates a real-world scenario where potential attackers have access to limited information about the system.

    \item \textit{Grey-box Scenario 2}: In this scenario, proposed in~\cite{efatinasab2024gangrid} for the GAN-GRID attack, the adversary does not have access to real data or the model architecture, but can query the model. The generator neural network is trained using reinforcement learning to produce data that the model classifies as stable. An example of this attack could occur by compromising a grid operator's dashboard through an insider attack~\cite{krause2021cybersecurity}, or leveraging malware like CRASHOVERRIDE~\cite{dragos2017crashoverride}, creating a backdoor for unauthorized access to these systems.  
    \end{itemize}

\section{Reference Attacks}
\label{sec:attacks}
In the white-box scenario, an attacker can exploit various state-of-the-art adversarial attacks.
While many such attacks have been proposed, most have been evaluated primarily in multi-class classification tasks and are not specifically tailored for binary classification problems like ours. We focus on a subset of attacks known for their effectiveness in revealing model vulnerabilities, particularly in decision-making contexts within smart grids, as supported by existing literature~\cite{efatinasab2024faultguard,efatinasab2024gangrid}.

In the grey-box scenarios, the same adversarial attacks are employed.
However, in the first grey-box scenario, adversaries only have access to genuine data and a surrogate model. They utilize this surrogate model to generate adversarial data, which is then tested against the primary model. This method assesses the robustness of the primary model without granting direct access to it. 

We will also consider another attack scenario (grey-box 2) called GAN-GRID~\cite{efatinasab2024gangrid}, a more sophisticated generative attack targeting the stability prediction system. This attack does not require access to the data or model architecture; it only needs the ability to query the model to craft adversarial inputs that can be classified as stable instances. These deceptive inputs could then be injected into the grid to mislead the stability prediction system.


A potential attacker's objective is to carry out the following attack:
\begin{align}\label{eq:attack}
\max_{\epsilon} & \quad L(f(x + \epsilon), y) &  & \text{s.t.} \; \|\epsilon\|_p \leq \gamma.
\end{align}
Equation~\ref{eq:attack} maximizes the loss \(L\) between the model's prediction \(f(x + \epsilon)\) and the true label \(y\) while constraining the perturbation \(\epsilon\) within a specified norm limit \(\|\epsilon\|_p \leq \gamma\).

The selected adversarial attacks for this study are as follows:
\begin{itemize}
  \item \textit{Fast Gradient Sign Method (FGSM):} FGSM efficiently generates adversarial examples using the sign of the gradient of the loss function and is widely used to benchmark the robustness of ML models~\cite{10510296}. 
  \item \textit{Basic Iterative Method (BIM):} BIM extends FGSM by iteratively applying small perturbations to input data. By gradually perturbing the input, BIM aims to enhance the potency of the attack and uncover vulnerabilities in ML models~\cite{10510296}.
  \item \textit{Randomized Fast Gradient Sign Method (RFGSM):} introduces randomness into FGSM iterations by incorporating random noise, enhancing attack diversity. Explores the impact of variability in adversarial perturbations, providing insights into model robustness against unpredictable attacks~\cite{tramèr2020ensemble}.
  
  \item \textit{Projected Gradient Descent (PGD):} PGD uses an iterative optimization approach like BIM, adding a projection step to keep perturbations within a predefined constraint set. This ensures perturbed examples remain within acceptable bounds, making PGD effective at crafting strong adversarial examples~\cite{10510296}. 

\item \textit{GAN-GRID}: is an adversarial attack that leverages the generator network of a GAN to create adversarial samples. The generator is trained using reinforcement learning methods, with a grid stability prediction system acting as a fixed discriminator or oracle. The training process focuses on refining the generator’s ability to produce effective adversarial samples based on feedback from the stability prediction system. In each training episode, the generator generates a sample from the latent space, which is then assessed by the stability prediction model. The model assigns a reward based on how closely the generated sample aligns with the target predictions, guiding the generator’s optimization process~\cite{efatinasab2024gangrid}.

\end{itemize}

\section{GAN-Stability: Our Proposed Stability Prediction System}
\label{methodology}
\begin{figure}[tbh]
    \centering
    \includegraphics[width=.8\columnwidth]{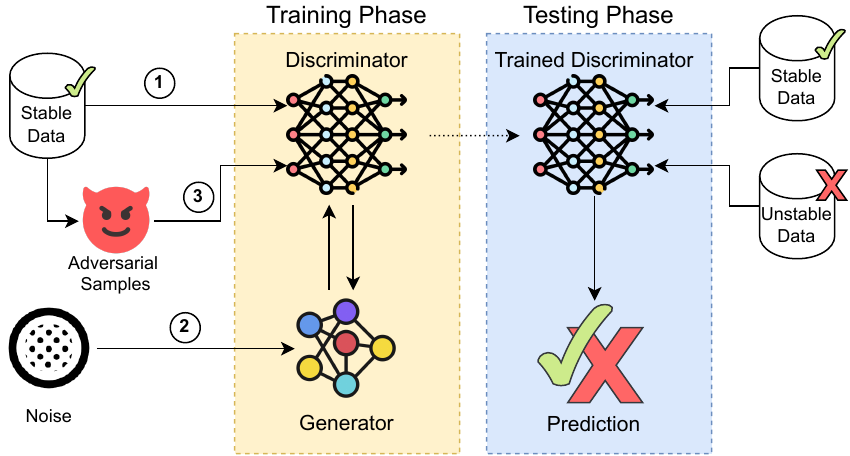}    
    \caption{GAN-Stability general architecture.}
    \label{fig:architecture}
\end{figure}

In this section, we present our proposed stability prediction system, whose architecture is summarized in Figure~\ref{fig:architecture}. In particular, training of the discriminator is done employing stable data (\circled{1}), synthetic Out-Of-Distribution (OOD) data relative to the stable class, generated by the generator from noise (\circled{2}), and adversarial samples generated by applying attacks to stable data (\circled{3}). 


\subsection{Architecture}\label{subsec:arch}
Introduced by Goodfellow et al.~\cite{goodfellow2020generative}, GANs consist of two neural networks---a generator and a discriminator---engaged in an adversarial game. 
The generator's objective is to produce synthetic data samples that closely resemble real ones, while the discriminator is trained to distinguish between genuine and fabricated samples. 


Our architecture employs a GAN-based framework for smart grid stability prediction, where the generator produces synthetic OOD samples relative to the stable class. These OOD samples may include measurement anomalies, fault-induced disturbances, instability events, or other deviations that do not conform to the stable data distribution.

The discriminator is trained to distinguish between real (stable) data and these synthetic OOD samples. By learning to differentiate stable conditions from any form of unstable behavior---without explicitly requiring labeled unstable data---our framework enhances anomaly detection and improves stability prediction. This approach enables the model to develop a more robust decision boundary, allowing for more accurate and generalized assessments of grid stability.

In this context, the generator does not have direct access to real data; its learning process relies solely on its interaction with the discriminator, which has access to both generated and real samples. 
The generator model, deliberately simpler than the discriminator, consists of four fully connected layers, with the number of neurons ranging from 12 (representing the number of features) to 128.
The discriminator, instead, comprises five fully connected layers, with the neuron count ranging from 12 to 512. 
The specific details of our GAN-based stability prediction system architecture can be seen in Table~\ref{tab:nn_architecture}.

\begin{table}[h]
    \centering
    \caption{The architecture of our GAN-based stability prediction system.}
    \begin{tabular}{c|c|c}
    \hline
        \textbf{Model} & \textbf{Architecture} & \textbf{Layers}\\
        \hline
        Generator  & feed-forward neural network & 4 (100, 128, 64,12)   \\
        Discriminator  & feed-forward neural network & 5 (160, 200, 256, 512,1)\\
        \hline
    \end{tabular}
    \label{tab:nn_architecture}
\end{table}

\subsection{Training}\label{subsec:training}
The training process in GANs is governed by a value function, $V(G, D)$, which accounts for both the generator \(G \) and discriminator \(D \). 
The training process involves solving 
\begin{equation}
\min \limits_{G} \max \limits_{D} V(G, D),
\end{equation}
where 
\begin{equation}
V(G, D) = \mathbb{E}_{p_{\text{data}}(x)}\log D(x) + \mathbb{E}_{p_{g}(x)}\log(1 - D(x)).
\end{equation}
 
The first term, \( \mathbb{E}_{p_{\text{data}}(x)}[\log D(x)] \), represents the expectation of the log-likelihood that the discriminator correctly identifies real data samples drawn from the distribution \( p_{\text{data}}(x) \). 
The second term, \( \mathbb{E}_{p_g(x)}[\log(1 - D(x))] \), corresponds to the expectation that the discriminator correctly identifies fake data samples generated by \( G \) from the generator's distribution \( p_g(x) \). 
The generator seeks to minimize this function by producing samples that the discriminator finds difficult to classify as fake, while the discriminator aims to maximize it by improving its ability to distinguish between real and generated data. 
This min-max game drives the adversarial training process, leading to improved generation of realistic data by \( G \) as training progresses.

During training, one model's parameter is updated while the others are kept fixed. 
Goodfellow et al.~\cite{goodfellow2020generative} demonstrate that when the generator is fixed, there exists a unique optimal discriminator $D^*(x)$ given by:
\begin{equation}
D^*(x) = \frac{{p_{\text{data}}(x)}}{{p_{\text{data}}(x) + p_{g}(x)}},
\end{equation}
which gives the probability that a sample $x$ belongs to the real data distribution $p_{\text{data}}$ rather than belonging to the generator's distribution $p_g$.
Additionally, they show that the generator $G$ is optimal when $p_{g}(x) = p_{\text{data}}(x)$, meaning the discriminator cannot distinguish between real and generated samples and assigns a probability of 0.5 to all samples, whether real or generated~\cite{8253599}.
In this standard formulation, the generator $G$ aims to minimize the objective by producing data that maximizes the discriminator's uncertainty (i.e., making $D(x)$ close to 0.5), while the discriminator $D$ aims to maximize its ability to correctly classify real versus generated data by minimizing this uncertainty.

In our approach, we employ a specialized training procedure for GAN models. The generator starts with random noise and aims to challenge the discriminator by producing synthetic data samples. However, rather than generating near-realistic data, the generator is specifically designed to create OOD samples relative to the stable class.

Meanwhile, the discriminator is trained on real data—comprising only stable instances from the dataset—alongside the synthetic OOD samples generated by the generator. Over time, the discriminator learns to differentiate between genuine stable data and these synthetic samples, effectively identifying deviations that fall outside the distribution of stable conditions. 

Unlike the traditional usage of GANs, where the goal is for the generator to eventually converge and produce data that closely resembles the real data such that $p_g(x) \approx p_{\text{data}}(x)$, we intentionally prevent the generator from reaching that stage of convergence. 
The optimization process for the generator is deliberately constrained to prevent it from fully converging. This modified min-max formulation aligns with our novel approach, where the generator’s objective is not to reach the typical GAN equilibrium but rather to help the discriminator learn more effective classification boundaries between stable data and OOD samples which could potentially involve unstable instances.

To encourage the generator to explore regions away from the stable distribution \( p_{\text{stable}} \), we add a regularization term to the generator's loss function. 
This term is referred to as the \textit{repulsion loss} and is defined as: 
\begin{equation}
\mathcal{L}_{\text{repulsion}} = \mathbb{E}_{x \sim p_g,\, s \sim p_{\text{stable}}} \left[ \text{ReLU}(m - \|x - s\|) \right],
\end{equation}
where:
\begin{equation}
\text{ReLU}(m - \|x - s\|) = 
\begin{cases} 
m - \|x - s\|, & \text{if } \|x - s\| < m, \\
0, & \text{if } \|x - s\| \geq m.
\end{cases}
\end{equation}

The generator minimizes the following objective:

\begin{equation}
L_G = -\mathbb{E}_{z \sim p_z} \left[ \log D(G(z)) \right] +  \mathbb{E}_{x \sim p_g,\, s \sim p_{\text{stable}}} \left[ \text{ReLU}(m - \|x - s\|) \right],
\end{equation}

In this context, \( x \) represents a batch of data generated by the generator, derived as \( x = G(z) \), where \( z \) is drawn from the latent distribution \( p_z \). The variable \( s \) denotes a batch of data from the stable distribution \( p_{\text{stable}} \), which represents regions in the data space the generator should avoid. The parameter \( m \) is a margin that determines the distance threshold for the repulsion effect.

Our objective is not for the generator to converge to the stable class but rather to guide it toward generating data that moves closer to the stable class distribution while maintaining a safe margin enforced by the repulsion loss. Since the latent distribution \( p_z \) (e.g., a Gaussian) has full support over \( \mathbb{R}^d \), the generator inherently possesses the capacity to map to any region in the data space \( X \). This means it is not restricted to reconstruct only the stable class but can generate samples in diverse and potentially unstable regions. To guide the generator away from the stable distribution \( p_{\text{stable}} \), we incorporate a repulsion loss that acts as a soft constraint. This loss creates an exclusion zone around the stable set \( S \subset X \), encouraging the generator to produce samples at least a margin \( m \) away from it. Formally, the repulsion loss drives generated samples into the set \( X_{\text{OOD}} \subset X \setminus B_m(S) \), defined as:
\[
X_{\text{OOD}} := \left\{ x \in X : \| x - s \| \geq m \text{ for all } s \in S \right\}\tag{8},
\]
where \( B_m(S) \) denotes the \( m \)-ball around the stable set. As training progresses, the distribution \( p_g \) of generated samples converges to high-density regions within this complement, effectively modeling plausible but non-stable areas in the data space. At the same time, adversarial pressure from the discriminator ensures these samples remain realistic, as it continues to evaluate whether \( G(z) \) resembles true data. This dynamic maintains the generator’s outputs on the data manifold while steering them outside the stable cluster, supporting the overall objective of improving discriminator boundaries between stable and OOD regions.

The choice of the margin m is grounded in the statistical characteristics of the stable class and the normalization applied to the dataset. After applying z-score normalization, each feature in the dataset has a standard deviation of approximately 1. Analysis of the stable class distribution showed that the majority of samples lie within three standard deviations from the mean , this follows the Empirical Rule, which states that approximately 99.7\% of values in a normal distribution fall within three standard deviations from the mean . To provide a conservative buffer and encourage exploration beyond this high-density region, we selected a margin of four standard deviations. This corresponds to \( m = 4 \) in the normalized space. This value was empirically validated to ensure that the generated samples \( p_g \) maintain a safe distance from the stable class \( p_{\text{stable}} \), avoiding overlap while still exploring plausible regions of the data space.


The discriminator’s ability to classify stable and unstable samples does not rely on the generator producing perfectly unstable real-world data. Instead, the generator acts as an adversarial tool, challenging the discriminator to refine its decision boundaries. Through this process, the discriminator learns to focus on key feature combinations that define stability, rather than relying on direct comparisons to real-world patterns. The generated OOD samples may encompass measurement anomalies, fault-induced disturbances, or other deviations that fall outside the stable data distribution. This approach enables the discriminator to generalize better, improving its ability to detect atypical scenarios without requiring explicit labels for unstable conditions.

Furthermore, we adopt a targeted strategy to extend the generator's training phase while allowing the discriminator to gain a prolonged upper hand. This approach diverges from conventional GAN setups that seek equilibrium between the generator and discriminator. Instead, our method intentionally introduces a controlled imbalance, enhancing the discriminator's capacity to classify generated data as unstable.
\begin{figure}[tbh]
    \centering
    \includegraphics[width=\columnwidth]{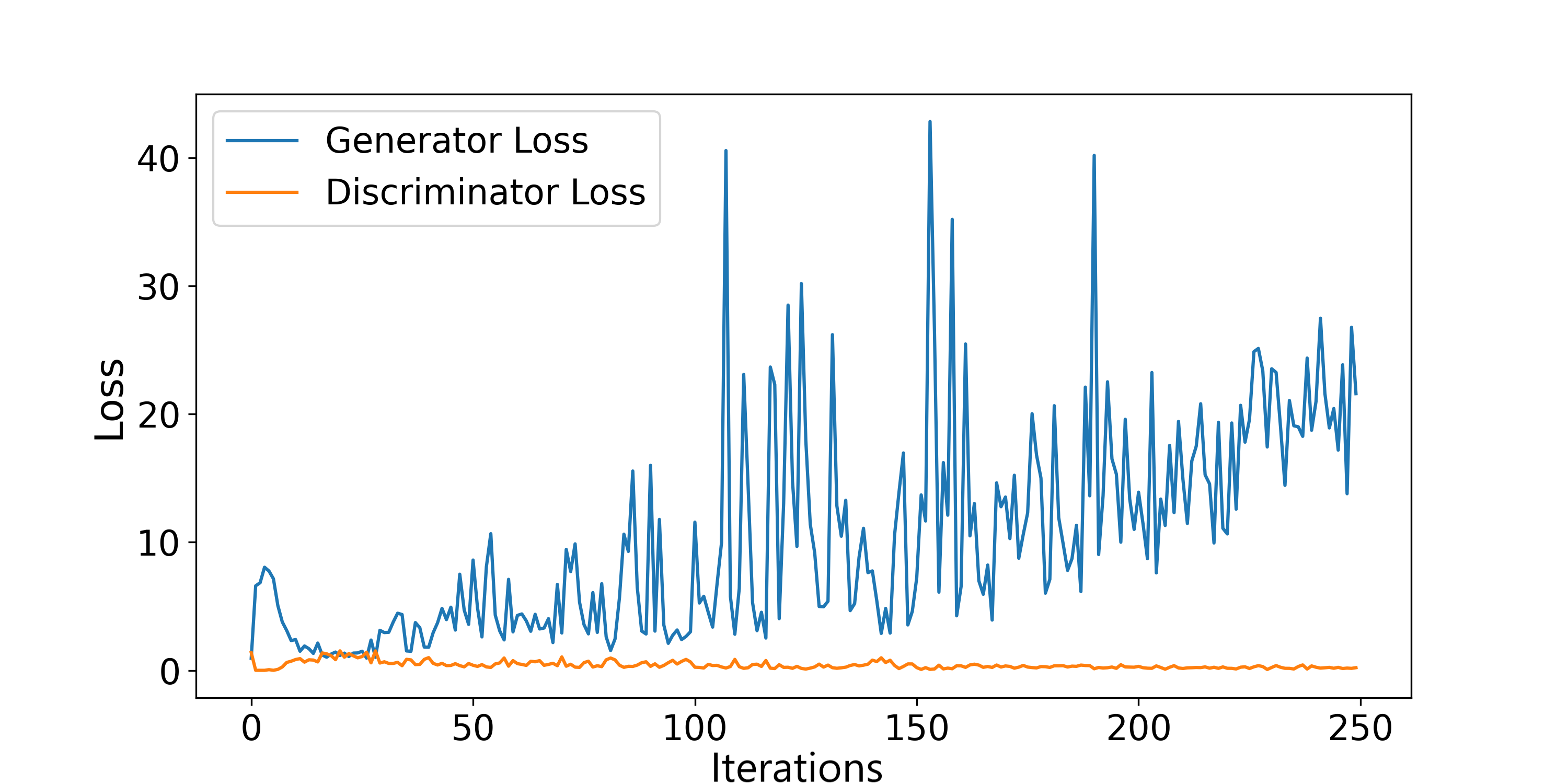}    
    \caption{Generator and Discriminator losses for the final model.}
    \label{fig:loss1}
\end{figure}

\begin{figure}[tbh]
    \centering
    \includegraphics[width=1\columnwidth]{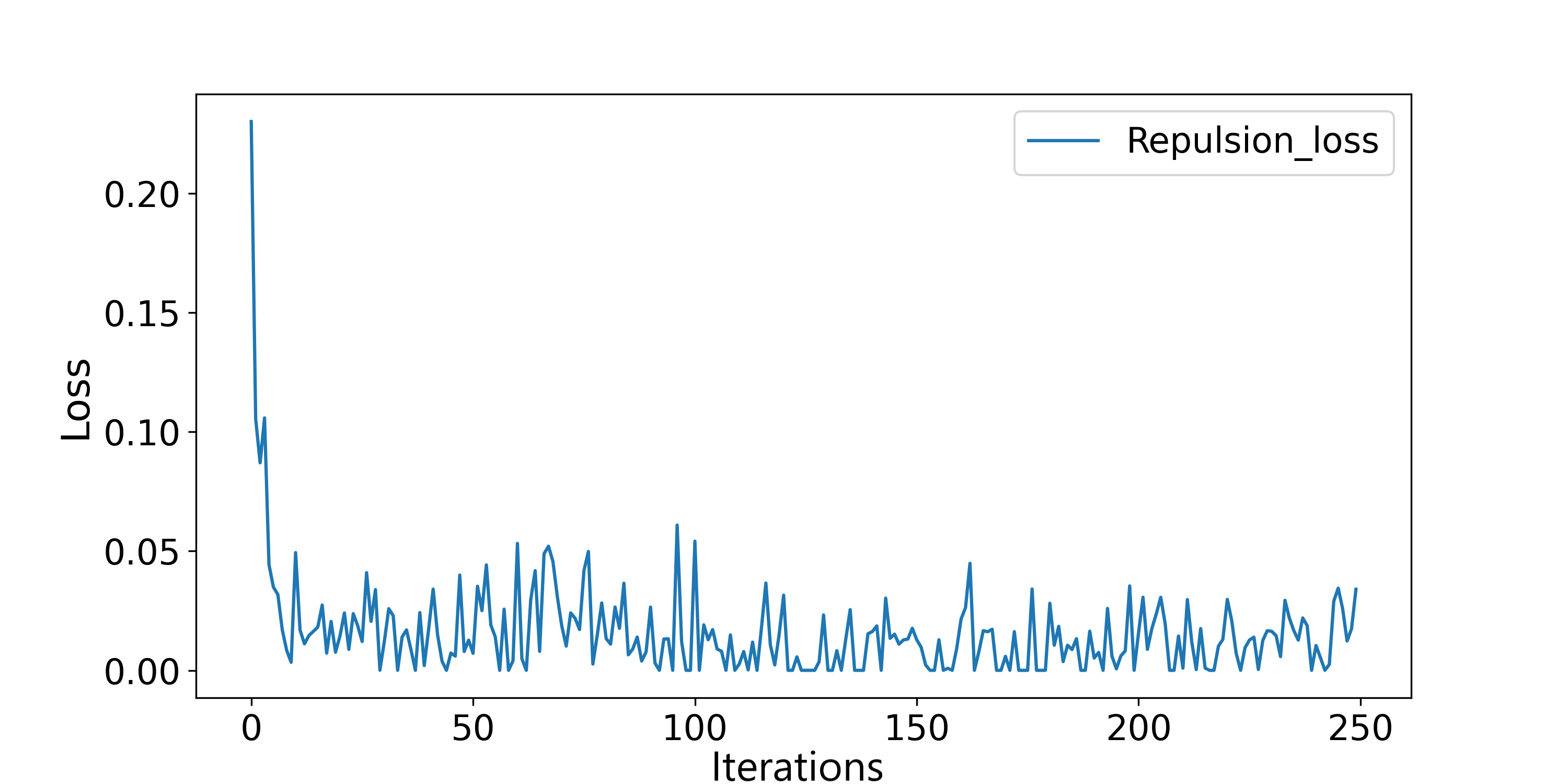}    
    \caption{Repulsion loss for the final model.}
    \label{fig:loss2}
\end{figure}
As shown in Figure~\ref{fig:loss1}, the generator’s loss starts at a high value and initially decreases as the generator begins to effectively challenge the discriminator. 
This early improvement highlights the generator's growing ability to deceive a discriminator that is still in the very early stages of learning, having been trained for less than 50 epochs. During this phase, the repulsion loss (Figure~\ref{fig:loss2}) starts with high values, reflecting the generator's proximity to the real data distribution. 
As training progresses, the repulsion loss decreases, indicating successful optimization that drives the generator to diverge from the stable data distribution. This process, coupled with the discriminator's growing advantage, results in a gradual increase in the generator's loss, eventually reaching values between 10 and 40.


The discriminator's architectural advantage, with its deeper network design, further reinforces this dynamic. Its capacity to learn complex representations allows it to maintain dominance throughout training, as seen by its relatively stable loss values in Figure~\ref{fig:loss1}. This structural superiority, coupled with prolonged training and repulsion loss, ensures the generator does not converge too closely to the stable data distribution. Instead, it generates OOD samples with respect to the stable class, which may be representative of any form of unstable behavior.

GAN-Stability is trained over 250 epochs with a learning rate of 0.0002 and a batch size of 4, ensuring thorough learning and refinement of the classification boundaries.

\subsection{Adversarial Training}
\label{adverserial_tr}


Together with real stable data (\circled{1} in Figure~\ref{fig:architecture})  and the samples crafted by the generator (\circled{2}), we introduce another source of data (\circled{3}) composed by a novel adversarial layer which is a novelty respect the traditional GAN training process.
The detailed steps about our proposed training approach can be found in Algorithm~\ref{algo}. Initially, the training steps follow the standard procedure as mentioned before, commencing with training the discriminator on real data labeled as stable (\circled{1}). Upon backpropagating the discriminator's loss, we proceed to generate fake data using the generator (\circled{2}),
This generation process begins by initializing with a random tensor of latent inputs, which the generator model then processes to generate samples. 
Subsequently, we evaluate these generated samples using the discriminator and backpropagate the loss accordingly.

Before advancing to the training of the generator, we introduce our novel layer of training. To enhance the adversarial detection capabilities of our discriminator, we introduce adversarial samples using the FGSM attack (\circled{3}). 
These samples are derived from the real data (in our case, the instances labeled as stable) that the discriminator was previously trained on and are generated through the exploitation discriminator model by the attack. The discriminator is then trained on these data with unstable labels, which can enhance its ability to detect FGSM samples as instances of instability within the grid.
However, it is important to note that the transferability property of these attacks enables the discriminator to also identify other types of attacks as instances of instability~\cite{10.1145/3607199.3607227}. At last, we continue with the training of the generator.

Although our primary task is stability prediction, categorizing these attacks as instances of instability serves as an additional warning mechanism for the grid operator. This classification provides an indication that a possible problem exists, where an intruder may be one of the contributing factors. 
\begin{algorithm}[H]
\caption{GAN Training with Adversarial Samples
}
\label{algo}
\For{$epoch = 1$ to $E=250$}{
    \textbf{Step 1 (real stable samples \circled{1}) } \\
    \For{each batch of real stable data $X_{\text{real}}$}{
        Compute discriminator output for real data: $D(X_{\text{real}})$ \\
        Compute discriminator loss on real data:  \\
        \[
        \mathcal{L}_{D\_real} = \mathbb{E}[\log D(X_{\text{real}})]
        \] \\
        Backpropagate $\mathcal{L}_{D\_real}$ to update $D$ \;
    }

    \textbf{Step 2 (generated unstable samples \circled{1})} \\
    \For{each batch of latent noise $z$}{
        Generate fake data: $X_{\text{fake}} = G(z)$ \\
        Compute discriminator output for fake data: $D(X_{\text{fake}})$ \\
        Compute discriminator loss for fake data: \\
        \[
        \mathcal{L}_{D\_fake} = \mathbb{E}[\log(1 - D(X_{\text{fake}}))]
        \] \\
        Backpropagate $\mathcal{L}_{D\_fake}$ to update $D$ \;
    }

    \textbf{Step 3 (adversarial samples \circled{3})} \\
    \For{each batch of real stable data $X_{\text{real}}$}{
        Generate adversarial samples using FGSM: \\
        $X_{\text{adv}} = X_{\text{real}} + \epsilon_{\text{FGSM}} \cdot \text{sign}(\nabla_{X} \mathcal{L}_{D\_real})$ \\
        Compute discriminator output for adversarial data: $D(X_{\text{adv}})$ \\
        Compute discriminator loss for adversarial data: \\
        \[
        \mathcal{L}_{D\_adv} = \mathbb{E}[\log(1 - D(X_{\text{adv}}))]
        \] \\
        Backpropagate $\mathcal{L}_{D\_adv}$ to update $D$ \;
    }

    \textbf{Step 4 (training the generator)} \\
    \For{each batch of latent noise $z$}{
        Generate fake data: $X_{\text{fake}} = G(z)$ \\
        Compute discriminator output for fake data: $D(X_{\text{fake}})$ \\
        Compute repulsion loss: \\
        \[
        \mathcal{L}_{\text{repulsion}} = \mathbb{E} \left[ \text{ReLU}(m - \|X_{\text{fake}} - X_{\text{real}}\|) \right]
        \] \\
        Compute generator loss with repulsion term: \\
        \[
        \mathcal{L}_{G} = \mathbb{E}[\log(D(X_{\text{fake}}))] +  \mathcal{L}_{\text{repulsion}}
        \] \\
        Backpropagate $\mathcal{L}_{G}$ to update $G$ \;
        }
}
\end{algorithm}

\section{Evaluation}
\label{Eval}
We now present the evaluation of GAN-Stability. As metrics, we use accuracy and F1 score to evaluate the models, defined as:

\begin{equation}
    Accuracy = \frac{TP + TN}{TP + FP + TN + FN},
\end{equation}

\begin{equation}
    F1 = \frac{2TP}{2TP + FP + FN},
\end{equation}

where $TP$ indicates the true positive, $TN$ the true negatives, $FP$ the false positives, and $FN$ the false negatives.

\subsection{Dataset}
\label{dataset}
The dataset used to evaluate our systems is an augmented version of the \emph{Electrical Grid Stability Simulated Dataset} from the University of California (UCI) Machine Learning Repository~\cite{misc_electrical_grid_stability_simulated_data__471}. 
The dataset is widely used for stability prediction~\cite{al2024smart, hangun2022forecasting, hangun2024quantum, raju2024smart, boutahir2023effective} and to test adversarial ML attacks against CPSs~\cite{mulo2023towards}. 
Initially, it consisted of 10,000 samples labeled as stable or unstable, representing simulation outcomes for a reference 4-node star mathematical model using DSGC concept. 
The model consists of two components: the first outlines the physical dynamics of electric power generation and its relationship with consumption loads, while the second defines an economic framework that links electricity prices to grid frequency as explained in \cite{schafer2015decentral,schafer2016taming}.
Through augmentation, the dataset has been expanded to 60,000 samples, capitalizing on the grid's inherent symmetry and resulting in a sixfold increase representing a permutation of the three consumers occupying three consumer nodes~\cite{breviglieri2023smartgrid}. With 12 primary predictive features and two dependent variables, the dataset provides valuable insights into grid stability dynamics.
Effective management of the dataset was achieved through a robust windowing technique, dividing it into predefined-size segments. Each window was iteratively created by traversing the data with a step size equal to half of the window size, set at 16 for our dataset.

Furthermore, leveraging our novel training technique that utilizes only one label from the dataset, we divided the dataset into two parts. The first part comprises all instances of the stable class (36.2\%), while the second part consists of the remaining instances of the unstable class (63.7\%) that we will employ for testing only. For the first part containing the stable class, we partitioned it into training (90\%) and test (10\%) subsets. Additionally, we reserved all instances of the unstable class for testing the GAN model. Thus, the testing dataset includes 10\% of the stable class instances and all instances of the unstable class. Our GAN model has access to only 32.85\% (90\% of the stable label) of the entire dataset for training, showcasing the most stringent data access constraints.


\subsection{Baseline Evaluation}
\label{baseline}
In the evaluation stage, we begin by establishing the baseline performance of our GAN-based stability prediction systems. This assessment is conducted before integrating any countermeasures, such as our novel adversarial training layer, and before exposing the system to adversarial attacks. Initially, we train our proposed model using the available training data (90\% of stable class with no samples from unstable class as discussed in Section~\ref{dataset}). Following the training phase, we proceed to evaluate the effectiveness of our GAN-based stability prediction system on the test dataset which compromises 10\% of stable instances and all samples from the unstable class. 

The results of our evaluation are noteworthy. Our GAN-based stability prediction system, even without access to the unstable class from the dataset during training, achieves a mean accuracy of 0.918. Specifically, when tested against stable instances of the test set, our model achieves an accuracy of 0.924, and when tested against unstable instances, it achieves an accuracy of 0.913. 


By allowing the discriminator to outperform the generator during training, adding the repulsion loss, and incorporating more layers and neurons, the adversarial process enhanced its ability to identify deviations from the stable class. This strong performance suggests that even without access to unstable class in the training, the diverse samples that the generator produces are sufficient for the discriminator to learn the complex characteristics of unstable behavior in any form.

\subsection{GAN-stability with Adversarial Training Evaluation}
In this section, we evaluate our GAN-based stability prediction system, augmented with our novel adversarial training layer as discussed in Section~\ref{adverserial_tr}. To assess the robustness of our model against adversarial attacks discussed in Section~\ref{sec:attacks}, we utilize the TorchAttacks~\cite{kim2020torchattacks} and Adversarial Robustness Toolbox (ART)~\cite{nicolae2018adversarial} library for attack implementation. We subject the discriminator model and testing dataset to various attacks, including FGSM, RFGSM, BIM, PGD and GAN-GRID, as detailed in Section~\ref{sec:attacks}. We employ an epsilon value of 0.05 for each attack, which signifies the strength of the attack and the magnitude of perturbation added to the data. The selection of 0.05 strikes a balance between the attack's power and the model's susceptibility, effectively challenging our model's ability to detect perturbations introduced by the attack. This value allows us to explore the attack's effectiveness while ensuring it remains within manageable bounds, thus facilitating a comprehensive evaluation of the model's robustness. By constraining the attack within manageable bounds, we mitigate potential risks such as susceptibility to ADS or expert human intervention in a real world scenario of a possible attack. In addition, we train the surrogate model, as previously described in Section~\ref{threat}, utilizing 70\% of the entire dataset (both labels) for training. 

Subsequently, we employ this surrogate model to generate adversarial samples through the mentioned attack techniques. Once the adversarial data is generated, we evaluate its efficacy against our primary GAN-based stability prediction system. This approach leverages the transferability of attacks. Despite operating within the constraints of a greybox scenario, where adversaries lack direct access to the primary model, this methodology enables us to assess the robustness of our system against potential real world adversarial threats. Following the generation of adversarial samples, we apply them to our GAN-based stability prediction model to induce misclassification and potentially inject fraudulent data into the grid. The objective is to deceive the stability prediction model into incorrectly classifying the situation as stable or unstable, rendering its output unreliable. A compromised stability prediction system can cause overvoltage, frequency deviations, and increased stress on grid components, leading to equipment failures, service disruptions, and reduced grid reliability~\cite{efatinasab2024gangrid}.

Our evaluation shows the effectiveness of our stability prediction system in accurately classifying state-of-the-art white-box adversarial attacks. With a mean accuracy of 0.989, our system adeptly identifies these attacks as belonging to the unstable class. Furthermore, in the grey-box setting, where attacks are launched against our stability prediction system using adversarial samples generated from a surrogate model, our system achieves a mean accuracy of 0.988. Furthermore, in the second grey-box scenario of the GAN-GRID attack, our framework achieves an accuracy of 0.991.  Detailed findings are presented in Table~\ref{tab:evaluation1}. To isolate the individual contributions of the adversarial training layer and the generator's OOD sample generation in enhancing the discriminator's robustness, we compare our full model against a baseline version without adversarial training. This baseline model, despite not being explicitly trained on adversarial examples, achieved a detection accuracy of 0.928 when evaluated under white-box adversarial attack conditions, 0.881 in the grey-box 1 scenario, and 0.958 in grey-box 2. These results demonstrate that the generator’s ability to produce realistic OOD samples significantly improves the model’s resilience, enabling it to distinguish between stable and unstable behaviors even without direct exposure to adversarial attacks.
However, incorporating the adversarial training layer further enhances performance across all scenarios.
By accurately identifying these attacks as non-stable instances, our stability prediction model can promptly raise an alarm to grid operators, enabling them to investigate the situation and take necessary precautions to prevent potential problems caused by the attack. 
This proactive approach reduces the need to employ another model acting as an ADS, which could introduce latency or overload to the grid communication system. 

        

\begin{table}[bthp]
    \centering
    \caption{Accuracy of GAN-Stability in classifying attacks as unstable samples. ``N/A'' indicates that the specific attack was not performed in that scenario.}
    \begin{tabular}{l|c|c|c|c|c}
        \hline
        \multirow{2}{*}{\textbf{Scenario}} & \multicolumn{5}{c}{\textbf{Accuracy}} \\ \cline{2-6}
        & FGSM & BIM & RFGSM & PGD & GAN-GRID \\
        \hline
        White-box  & $1.000$ & $0.957$ & $1.000$ & $1.000$ & N/A \\
        Grey-box 1 & $0.987$ & $0.988$ & $0.987$ & $0.989$ & N/A \\
        Grey-box 2 & N/A & N/A & N/A & N/A & $0.991$ \\
        \hline
    \end{tabular}
    \label{tab:evaluation1}
\end{table}

\begin{figure}[tbh]
    \centering
    \includegraphics[width=1\columnwidth]{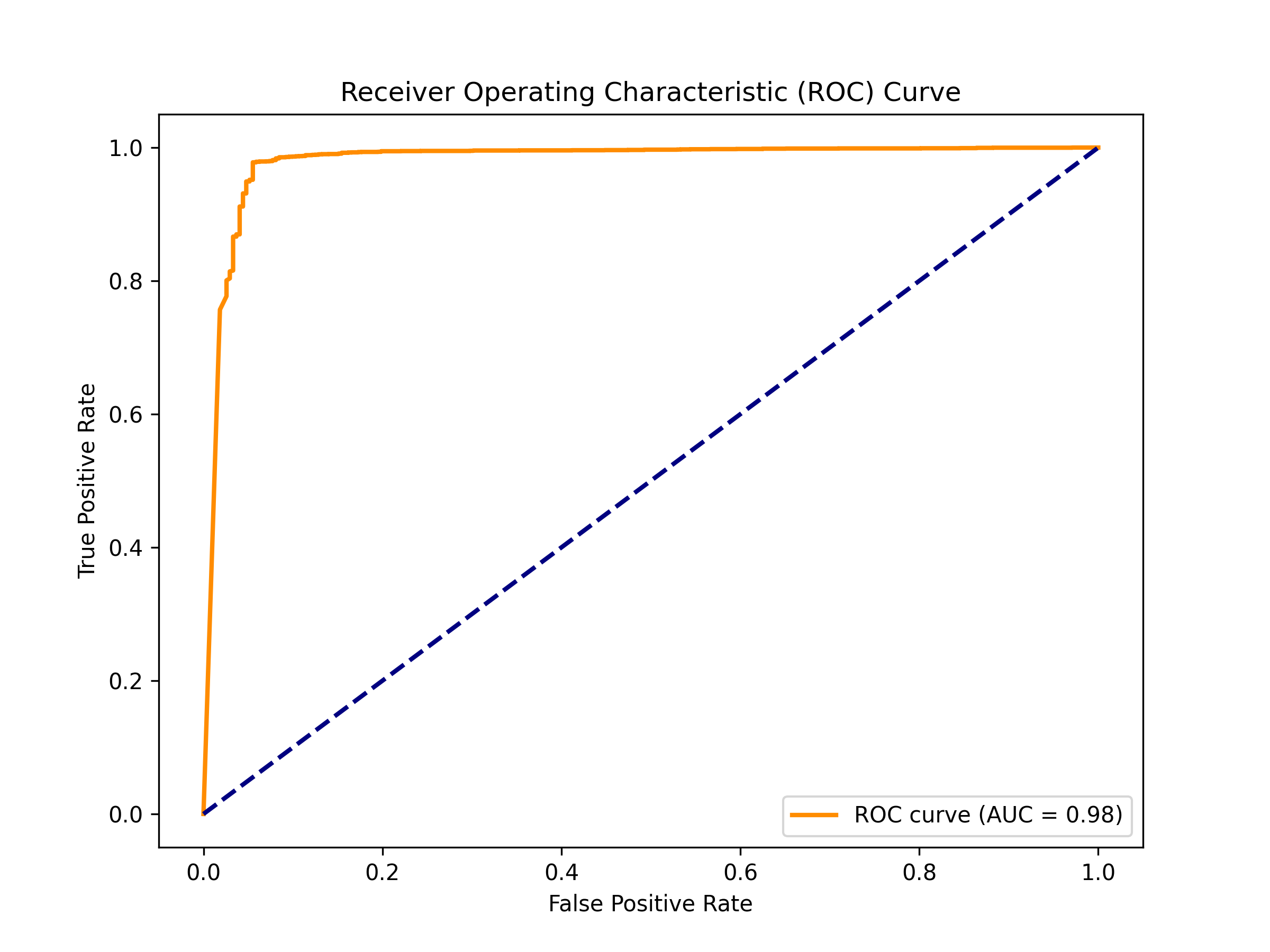}    
    \caption{ROC curve for classification performance of GAN-Stability augmented with the adversarial training layer}
    \label{fig:roc_curve}
\end{figure}

Following the augmentation of GAN-Stability 
with the new adversarial training layer, it becomes essential to re-evaluate its performance against the main task of stability prediction to ensure that the augmentation does not compromise the efficacy of the system. In these settings, our system achieves a mean accuracy of 0.981 considering both classes.
Specifically, when tested against stable instances of the test set, our model achieves an accuracy of 0.911, and when tested against unstable instances, it achieves an accuracy of 0.986. 
The integration of the adversarial attack layer significantly boosts the performance of the discriminator, despite the small deviations introduced by the attack. By incorporating adversarial samples, the discriminator is exposed to subtle perturbations of real stable data. These adversarial examples act as challenging cases, forcing the discriminator to become more sensitive to nuanced changes in the input data. Although these perturbations may seem minor, they encourage the discriminator to refine its decision boundary and strengthen its ability to differentiate between truly stable and unstable instances. 
The combined impact of the repulsion loss, the discriminator's architectural superiority, the extended training period, and the adversarial training layer ensures the system is equipped to accurately identify instability while maintaining strong generalization across diverse data distributions.
The summary of the results is presented in Table~\ref{tab:eval2}.
Also, as shown in Figure~\ref{fig:roc_curve}, the ROC curve illustrates the GAN-Stability's performance across different threshold values.  

\begin{table}[!htpb]
 \centering
    \caption{Accuracy and F1 score of GAN-Stability with and without Adversarial Training (AT).}
    \begin{tabular}{l|c|c|c}
        \hline
        \textbf{Task} & \textbf{AT} & \textbf{Accuracy} & \textbf{F1 Score} \\
        \hline
        \multirow{2}{*}{Stable class} & \xmark & 0.924 & - \\
        & \cmark & 0.911 & - \\ \hline
        \multirow{2}{*}{Unstable class} & \xmark & 0.913 & - \\
        & \cmark & 0.986 & - \\
        \hline
         \multirow{2}{*}{Both Classes} & \xmark & 0.918 & 0.955 \\
        & \cmark & 0.981 & 0.99 \\
        \hline
    \end{tabular}
    \label{tab:eval2}
\end{table}

\subsection{Benchmarking Against State-of-the-Art}
In this section, we compare our proposed system with state-of-the-art models from the literature using the same dataset. As shown in Table~\ref{tab1}, many existing models achieve high accuracy. However, these models are trained on the entire dataset with access to both labels, an approach that may not be feasible in real-world scenarios due to the significant time, resources, and expertise required to collect instability instances. Notably, training such models with only one label is nearly impossible, underscoring a key advantage of our method.

Despite being trained without unstable data, our GAN-based approach exhibits superior robustness compared to some supervised models. The stability of the GAN framework facilitates effective learning without explicit exposure to unstable samples, while the generator produces OOD samples that serve as adversarial challenges, improving the discriminator’s generalization beyond the training distribution. Although certain supervised models achieve higher accuracy, they rely on labeled unstable data, whereas our model is tested on a significantly larger and more diverse dataset that includes all unstable cases.
Furthermore, existing models do not address the risk of adversarial attacks or incorporate countermeasures to mitigate them. In contrast, our approach not only operates with a single-label training paradigm, making it more practical for real-world deployment, but also integrates adversarial training layers, enhancing resilience against such threats—an aspect often overlooked in prior research.

\begin{table}[htbp]
\caption{Comparison of GAN-Stability with other state-of-the-art models. Only our system evaluates adversarial robustness, which is not reported in prior works. ``N/A'' indicates that adversarial robustness was not assessed in those studies.}
\centering
\begin{threeparttable}
\begin{tabular}{l|c|c|c|c}
\hline
\textbf{Model} & \makecell{\textbf{Stable}\\\textbf{Access}} & \makecell{\textbf{Unstable}\\\textbf{Access}} & \textbf{Accuracy} & \makecell{\textbf{Adversarial}\\ \textbf{Robustness}} \\
\hline
ANN~\cite{breviglieri2021predicting} & \cmark & \cmark & 0.996 & N/A 
\\
ANN~\cite{sym15020289} & \cmark & \cmark & 0.985 & N/A 
\\
XGBoost~\cite{efatinasab2024gangrid} & \cmark & \cmark & 0.994 & N/A\\
CatBoost~\cite{ALLAL2024108304} & \cmark & \cmark & 0.996 & N/A \\ 
CNN~\cite{MOSTAFA2022100363} & \cmark & \cmark & 0.870 & N/A \\
LSTM~\cite{9079864} & \cmark & \cmark & 0.990 & N/A \\
ANN~\cite{mohsen2023efficient} & \cmark & \cmark & 0.973 & N/A \\
Decision Tree~\cite{efatinasab2024gangrid} & \cmark & \cmark & 0.974 & N/A \\
KNN~\cite{efatinasab2024gangrid} & \cmark & \cmark & 0.875 & N/A \\
\textbf{GAN-Stability} & \cmark & \xmark & \textbf{0.981} & \textbf{0.989}\tnote{1} - \textbf{0.988}\tnote{2} \\
\hline
\end{tabular}
\begin{tablenotes}\footnotesize
    \item[1] Adversarial robustness measured under white-box conditions.
    \item[2] Adversarial robustness measured under grey-box conditions.
\end{tablenotes}
\label{tab1}
\end{threeparttable}
\end{table}

\subsection{Training Time}
The experiments in this paper were conducted on Kaggle, utilizing a free cloud-based resource with the following specifications: Intel(R) Xeon(R) CPU @ 2.20GHz, 32 GB of RAM, running Linux Ubuntu, and equipped with Python 3.10.14. 
In this section, we present the training time per epoch of our framework, both with and without the inclusion of the new adversarial training layer. The results show significant differences in training time between our GAN model configurations with and without the new adversarial training layer. Training without this layer takes approximately \(8 \pm 2\) seconds per epoch, while incorporating the layer increases this to \(14 \pm 2.5\) seconds per epoch. Although the new adversarial training layer adds considerable computational overhead, it is essential to consider the trade-offs involved. The enhanced system security and potential performance improvements justify the increased training time, highlighting the benefits of integrating the new adversarial training layer into the framework.

\subsection{Hardware Implementation}
In order to demonstrate the practicality of our framework in real-world applications, we conducted an experiment using a Raspberry Pi 4 Model B, which features a 4GB RAM and a Quad-core Cortex-A72 (ARM v8) 64-bit SoC running at 1.5GHz, Raspberry Pi OS 12 (Debian Bookworm porting)~\cite{rpi}. We used Python 3.11.2 and Pytorch 2.4.1. This configuration represents one of the most basic and cost-effective hardware setups available on the market. The complete results of our experiment can be found in Table~\ref{tab:rpi}.

\begin{table}[htb]
\centering
\def\arraystretch{1.1}
\caption{Time results (in seconds) for training of one epoch and testing for one batch for our models in a Raspberry Pi.}
\label{tab:rpi}
\begin{tabular}{l|c|c}
\hline
  \textbf{Model} & \textbf{Train} & \textbf{Test} \\ 
\hline
GAN-Stability & $40.61 \pm 1.93$ & $0.00690 \pm  0.0047$ \\
GAN-Stability (with adv. training) & $58.91 \pm 2.08$ & $0.00646 \pm 0.0030$ \\
\hline
\end{tabular}
\end{table}

Despite the inherent limitations of this hardware, our results indicate that the training time remains manageable even with the integration of our new adversarial training layer. Specifically, the model was able to train within reasonable time frames, making local training feasible without the need for high-end computational resources. Each epoch took approximately $58.91 \pm 2.08$ seconds when the adversarial training layer was activated, compared to $40.61 \pm 1.93$ seconds per epoch recorded without this layer. While the increase in time is notable, it remains acceptable given the enhanced security benefits and potential performance gains provided by the adversarial layer.
Additionally, the testing time for each batch of data was negligible, ranging between \(0.00646\) and \(0.00690\), or approximately $6.5$ to $7$ milliseconds, suggesting that the model can effectively perform real-time decision-making tasks—a critical requirement in time-sensitive environments such as smart grids. 

The Raspberry Pi’s ability to handle these tasks underscores the lightweight nature of our framework, demonstrating that it can function efficiently even on constrained hardware. While the Raspberry Pi serves as a basic testbed, smart grid implementations could leverage more powerful hardware, such as FPGAs or GPUs, for local training or utilize cloud-based resources for more demanding tasks, especially when dealing with large datasets.



\section{Limitation}
\label{limit}

In this section, we outline some of the limitations of this study.

\begin{itemize}
\item \textbf{Real-World Deployment Constraints:} While the model demonstrated efficient real-time decision-making on a single-board computer, real-world smart grid environments may introduce additional challenges, such as communication latency, hardware limitations, and integration complexities. Deploying the proposed model requires seamless compatibility with existing grid monitoring, control, and communication systems, including Supervisory Control and Data Acquisition (SCADA) systems, phasor measurement units (PMUs), and energy management systems (EMS). Achieving this integration may necessitate modifications or middleware solutions.

\item \textbf{Lack of Explicit Interpretability:} GAN-based models are inherently complex and may lack transparency, making it difficult for operators to understand the reasoning behind specific instability detections.

\item \textbf{Requirement for a Comprehensive Stable-Class Dataset:} The effectiveness of this approach depends on access to a diverse and comprehensive dataset of stable-class instances. If the dataset lacks sufficient variability, the model may struggle to distinguish true instability from rare but valid variations within stable conditions.

\item \textbf{Evaluation on Larger Datasets:} While the model has been tested on a dataset containing 60000 stable and unstable samples, its performance on significantly larger datasets with greater variability due
to the lack of publicly available large-scale smart grid
dataset remains untested. Further validation on diverse grid configurations, renewable energy penetration levels, is necessary to fully assess its scalability and robustness.
\end{itemize}

\section{Conclusion}
\label{conclusion}
Our paper presents a pioneering framework using a GAN model to predict smart grid stability, effectively addressing the challenge of limited data accessibility by focusing on stable instances from available datasets. This approach aligns with real-world constraints where instability data is scarce. Our model achieves a commendable accuracy of 0.981 in stability prediction, even without instability instances in the training data, using just 32.85\% of the dataset.
Additionally, by incorporating a novel adversarial training layer, our framework demonstrates robustness against state-of-the-art adversarial attacks in both whitebox and greybox scenarios, classifying these attacks as instances of instability. 
We benchmarked our model against state-of-the-art approaches and found that while it may not surpass current best performances, it still offers high accuracy and enhanced robustness, a feature not commonly found in existing literature.
In summary, our study advances stability prediction for smart grids by offering high accuracy and resilience, addressing data scarcity challenges, and enhancing operational efficiency and reliability in evolving energy landscapes with emerging cybersecurity threats.

\subsection{Future Work} \label{future_work} While our framework demonstrates strong performance in stability prediction and adversarial robustness, several directions remain for future exploration. First, we plan to expand the model’s capabilities by incorporating real-world instability instances as they become available. This would enable direct training on both stable and unstable samples, potentially through continual learning techniques. Second, we seek to enhance the diversity and semantic quality of the generated OOD samples. Approaches may include conditioning the generator on domain-specific attributes or employing contrastive objectives to promote clearer separation between stable and unstable regions in the latent space. Another promising direction involves integrating domain knowledge and physics-informed constraints into the model to ensure interpretability and alignment with the underlying dynamics of power grids. Finally, we are exploring deployment in real-time grid monitoring environments. This will require adapting the model to meet strict latency constraints, process streaming data, and maintain robustness over extended operational periods without retraining. Additionally, we are considering evaluation with historical SCADA/PMU data to assess the feasibility of a real-world application.

\bibliographystyle{elsarticle-num} 
\bibliography{bare_jrnl}

\end{document}